# A novel concept of pseudo ternary diffusion couple for the estimation of diffusion coefficients in multicomponent systems

Neelamegan Esakkiraja and Aloke Paul*

Department of Materials Engineering, Indian Institute of Science, Bangalore 560012, India

*Corresponding author: E–mail: aloke@iisc.ac.in

Tel.: 918022933242, Fax: 91802360 0472

**Abstract**

A pseudo ternary diffusion couple technique in a multicomponent system by simplifying the mathematical complications of Onsager formalism is proposed for the estimation of composition dependent values of the interdiffusion coefficients. This is otherwise impossible following the conventional method in a system with more than three components. Other alternative methods estimate the average diffusion coefficients over a composition range of random choice and lack physical significance. This method can be followed in a multicomponent system with any number of components on the condition that only three components develop diffusion profiles keeping others as constant.

Diffusion studies in inhomogeneous multicomponent systems represent an unsolved challenge. Mathematical complexities based on Onsager's formalism [1-3] makes it impossible to estimate meaningful diffusion coefficients [4, 5]. Due to relative ease, most of the experimental studies in the academic world are conducted in binary systems for an understanding of the basic atomic and phenomenological diffusion mechanisms. The studies in ternary systems are much smaller in number because of increase in complexities; however, these are used to understand diffusional interactions among components. No experimental studies are available estimating the composition dependent values of the diffusion coefficients following the conventional method in a system with a higher number of components. This is simply not possible because of mathematical complexities [4, 5]. On the other hand, to achieve a property balance, most of the material systems in applications are multicomponent. Diffusion coefficients are important to understand microstructural evolution and many physicomechanical properties of a material. Therefore, this difficulty brings an unbridgeable gap between academics studying basics of diffusion mechanisms in simple systems and industries developing complicated material systems for various applications.

To counter the mathematical and experimental complications in ternary and multicomponent systems [4], few alternative experimental methods were developed with mixed success [5]. Thompson and Morral [6] developed the concept of square root diffusivity for the estimation of constant diffusion coefficient, which could not be used for composition range more than 5 at.% [7]. Dayananda and Sohn [8] developed a method for the estimation of average interdiffusion coefficients. However, these are not material constants but depend on the composition range of the diffusion couple and the composition range over which these average values are estimated. Therefore, these data lack physical significance compared to the composition dependent values of the diffusion coefficients estimated following the conventional method [5]. At present few other numerical methods are being proposed in multicomponent systems [9, 10]; however, further refinement and rigorous analyses in many systems with different situations might be required before adopting them with a certain degree of confidence. Most importantly, there is a need for a purely experimental method with straightforward and easy to follow steps such that the research students and engineers can adopt in their studies without much difficulty for the estimation of meaningful composition dependent (not average) diffusion coefficients.

Fulfilling this requirement, Paul and co-workers recently proposed a pseudo binary approach [11,12]. This needs only a single diffusion couple in a multicomponent system with an added advantage of estimation of composition-dependent (not average) values of the interdiffusion coefficients and even the intrinsic diffusion coefficients [12], which was not possible by the conventional or any other alternative methods even in a ternary system [5]. This method is established based on the concept that only two components develop diffusion profiles in a ternary or multicomponent system countering the mathematical complications of the Onsager formalism. This is already used successfully in few systems [12,13,14,15]. It should be noted here that this is named differently by different groups as pseudo binary or quasi binary although the basic concept is the same irrespective of the name used. Most importantly, it gives a unique opportunity to relate diffusion rates of components with defects and thermodynamic driving forces for understanding the growth kinetics and microstructural evolution in a multicomponent system of practical importance [16,17]. However, this approach has a drawback that the contribution of cross diffusivity terms towards the interdiffusion flux i.e. diffusional interactions among different components cannot be determined, which sometimes play a major role on phenomenological diffusion process.

Therefore, in this manuscript, we propose an experimental pseudo ternary approach simplifying the mathematical complications of the Onsager formalism such that only three components develop diffusion profiles keeping the composition of all other components as constant in a multicomponent system. This can be considered as the extension of the pseudo binary approach [11, 12]. As explained, we can now estimate the composition dependent (not average) values of the main and cross interdiffusion coefficients in a system with more than three components, which is not possible following the conventional or any other available alternative methods.

For the sake of explanation of the experimental method and subsequent analysis, we consider a quaternary Ni-Co-Fe-Mo system although this can be used in a system with even higher number of components without any further modification of the concept. Following the Onsager's formalism, the interdiffusion fluxes ($\widetilde{J}_i$) of different components (*i*) with respect to the interdiffusion coefficients are related by [1, 4, 5]

$$\widetilde{J}_i = -\sum_{j=1}^{n-1} \widetilde{D}_{ij}^n \frac{dC_j}{dx} = -\sum_{j=1}^{n-1} \widetilde{D}_{ij}^n \frac{1}{V_m} \frac{dN_j}{dx}, \qquad (1)$$

$\frac{dC}{dx} = \frac{1}{V_m} \frac{dN}{dx}$ is the concentration gradient considering a constant molar volume since the composition dependent lattice parameter variations are not known in a multicomponent system. In such a situation, the interdiffusion fluxes estimated with respect to different components are related by [4]

$$\sum_{i=1}^{n} \widetilde{J}_i = 0 \qquad (2)$$

$n^{th}$ component is considered as the dependent variable. Considering the major component Ni as the dependent variable, the interdiffusion fluxes in the Ni-Co-Fe-Mo system can be written as

$$\widetilde{J}_{Co} = -\widetilde{D}_{CoCo}^{Ni} \frac{1}{V_m} \frac{dN_{Co}}{dx} - \widetilde{D}_{CoFe}^{Ni} \frac{1}{V_m} \frac{dN_{Fe}}{dx} - \widetilde{D}_{CoMo}^{Ni} \frac{1}{V_m} \frac{dN_{Mo}}{dx} \qquad (3a)$$

$$\widetilde{J}_{Fe} = -\widetilde{D}_{FeFe}^{Ni} \frac{1}{V_m} \frac{dN_{Fe}}{dx} - \widetilde{D}_{FeCo}^{Ni} \frac{1}{V_m} \frac{dN_{Co}}{dx} - \widetilde{D}_{FeMo}^{Ni} \frac{1}{V_m} \frac{dN_{Mo}}{dx} \qquad (3b)$$

$$\widetilde{J}_{Mo} = -\widetilde{D}_{MoMo}^{Ni}\frac{1}{V_m}\frac{dN_{Mo}}{dx} - \widetilde{D}_{MoCo}^{Ni}\frac{1}{V_m}\frac{dN_{Co}}{dx} - \widetilde{D}_{MoFe}^{Ni}\frac{1}{V_m}\frac{dN_{Fe}}{dx} \tag{3c}$$

$$\widetilde{J}_{Ni} + \widetilde{J}_{Co} + \widetilde{J}_{Fe} + \widetilde{J}_{Mo} = 0 \tag{3d}$$

$\widetilde{D}_{ii}^n$ and $\widetilde{D}_{ij}^n$ are the main and cross interdiffusion coefficients and $n$ (Ni in this set of equations) is the dependent variable [4]. Therefore, $(n-1)^2$ interdiffusion coefficients are required to determine in $n$ component system. The interdiffusion flux of the component considered as the dependent variable is related to the interdiffusion fluxes of other components following Equation 3d. [1, 5]. Interdiffusion fluxes can be calculated utilizing the Wagner or den Broeder's relation for constant molar volume as [4, 5]

$$\widetilde{J}_i = -\frac{N_i^+ - N_i^-}{2tV_m}\left[(1-Y_i^*)\int_{x^{-\infty}}^{x^*}Y_i dx + Y_i^*\int_{x^*}^{x^{+\infty}}(1-Y_i)dx\right] \tag{4}$$

The composition normalized variable is expressed as $Y_i = \frac{N_i - N_i^-}{N_i^+ - N_i^-}$. $N_i^-$ and $N_i^+$ are the compositions in mole (or atomic) fraction in left ($x^{-\infty}$) and right hand ($x^{+\infty}$) side of the unaffected parts of the diffusion couple. It can be seen in Equation 3 that in a quaternary system, total nine interdiffusion coefficients should be estimated. On the other hand, only three independent interdiffusion fluxes can be estimated in a particular diffusion couple. Therefore, following this conventional method, we need three diffusion couples such that all of them intersect at one particular composition at which these data can be estimated since the diffusion coefficients are material constants which vary with composition because of change in thermodynamic driving forces and defects assisting the diffusion process. It is practically impossible to predict the end member compositions of three different diffusion couples in a four-component space such that all of them intersect at one particular composition because of different serpentine diffusion (composition) paths followed by individual diffusion couples [4, 18]. Therefore, no experiments are available in a quaternary or higher order systems following the conventional method estimating the composition dependent (not average over a composition range) values of the interdiffusion coefficients. Because of this reason only, an alternative method was developed by Dayananda and Sohn [8] for the estimation of average interdiffusion coefficients. The major advantage of this approach is that only one diffusion couple is required in a system with any number of components. However, these are average over a composition range of random choice leading to different values for different composition ranges. Even the values will be different when a common composition range is considered but from diffusion couples with different composition ranges of the end members. This additional complication comes from the fact that it is not easy to control the diffusion paths which depends on the initial end member alloys [19]. Therefore, the estimated data are vague without any physical significance [5]. This is still practiced to get a rough idea of the diffusion coefficients since no other efficient and straightforward experimental method is available in multicomponent systems.

Because of the difficulties explained above, we need an approach for the estimation of composition dependent values of the interdiffusion coefficients. To circumvent the mathematical complications based on Onsager formalisms [2, 3], we propose a pseudo ternary approach. Therefore, in the Ni-Co-Fe-Mo quaternary system used to demonstrate this approach, we make one of the components, for example, Mo to remain constant without producing any diffusion profiles. It further means that we keep the content of Mo as same in both the end members of a diffusion couple with the expectation that it remains constant throughout the diffusion couple. Experimental results of two such diffusion couples are shown in Figure 1(a) (DF1: $Ni_{95}Mo_5/Ni_{75}Co_{10}Fe_{10}Mo_5$) and (b) (DF2: $Ni_{90}Fe_5Mo_5/Ni_{85}Co_{10}Mo_5$) which are used for further analysis. The method of producing diffusion couples can be found in Chapter 3 in Ref. [5] in details. Experimental procedure for this study is described in the supplementary file.

In such a situation, $\widetilde{J}_{Mo}$ and $\frac{dN_{Mo}}{dx}$ are equal to zero so that series of Equation 3 reduces to

$$\widetilde{J}_{Co} = -\widetilde{D}_{CoCo}^{Ni} \frac{1}{V_m} \frac{dN_{Co}}{dx} - \widetilde{D}_{CoFe}^{Ni} \frac{1}{V_m} \frac{dN_{Fe}}{dx} \tag{5a}$$

$$\widetilde{J}_{Fe} = -\widetilde{D}_{FeFe}^{Ni} \frac{1}{V_m} \frac{dN_{Fe}}{dx} - \widetilde{D}_{FeCo}^{Ni} \frac{1}{V_m} \frac{dN_{Co}}{dx} \tag{5b}$$

$$\widetilde{J}_{Ni} + \widetilde{J}_{Co} + \widetilde{J}_{Fe} = 0 \tag{5c}$$

Therefore, the relations above reduces similar to a ternary system in which only four interdiffusion coefficients are required to estimate. This can be easily achieved by making only two diffusion couples and by forcing them to intersect at one particular composition for estimation of the diffusion coefficients at that composition. The same concept can be followed in a system with even higher number of components by preparing the diffusion couples such that only three components develop the diffusion profiles keeping all other components constant.

Although the interdiffusion flux and the composition gradient are zero, Mo is present in the system such that

$$N_{Ni} + N_{Co} + N_{Fe} + N_{Mo} = 1 \tag{6a}$$

Therefore, the compositions of the components which develop the diffusion profiles are required to modify as

$$M_{Ni} + M_{Co} + M_{Fe} = 1 \tag{6b}$$

where $M_{Ni} = \dfrac{N_{Ni}}{1 - N_{Mo}}$, $M_{Co} = \dfrac{N_{Co}}{1 - N_{Mo}}$, $M_{Fe} = \dfrac{N_{Fe}}{1 - N_{Mo}}$ since $N_{Mo} = 0.05$ has a fixed value.

One should be very careful to normalize the profiles correctly. This method explained above can be followed in a solid solution in which components can occupy any lattice position randomly. In intermetallic compounds, the components should be coupled depending on the sublattices they occupy.

Following, the actual composition profiles ($N_i$ vs. $x$) of the two diffusion couples used for the estimation of the diffusion data should be re-plotted with respect to the modified composition profiles ($M_i$ vs. $x$), as shown in Figure 1(c) and (d). These are then plotted on Gibb's triangle to find the composition of the intersection, as shown in Figure 2. The composition of intersection is $M_{Ni} = 0.930$, $M_{Co} = 0.031$, $M_{Fe} = 0.039$. Therefore, the actual quaternary composition at which the interdiffusion coefficients are estimated can be recalculated following Equation 6 as $Ni_{0.883}Co_{0.030}Fe_{0.037}Mo_{0.05}$. The relations for the estimation of the interdiffusion coefficients and the interdiffusion fluxes with respect to the modified compositions can be expressed as

$$\tilde{J}_{Co} = -\tilde{D}_{CoCo}^{Ni} \frac{1}{V_m} \frac{dM_{Co}}{dx} - \tilde{D}_{CoFe}^{Ni} \frac{1}{V_m} \frac{dM_{Fe}}{dx} \tag{7a}$$

$$\widetilde{J}_{Fe} = -\widetilde{D}^{Ni}_{FeFe}\frac{1}{V_m}\frac{dM_{Fe}}{dx} - \widetilde{D}^{Ni}_{FeCo}\frac{1}{V_m}\frac{dM_{Co}}{dx} \tag{7b}$$

$$\widetilde{J}_i = -\frac{M_i^+ - M_i^-}{2tV_m}\left[(1-Y^*_{M_i})\int_{x^{-\infty}}^{x^*} Y_{M_i}dx + Y^*_{M_i}\int_{x^*}^{x^{+\infty}}(1-Y_{M_i})dx\right] \tag{7c}$$

where $Y_{M_i} = \frac{M_i - M_i^-}{M_i^+ - M_i^-}$. $M_i^-$ and $M_i^+$ are the modified compositions in mole (or atomic) fraction in left ($x^{-\infty}$) and right hand ($x^{+\infty}$) side of the unaffected parts of the diffusion couple The experimentally measured composition profiles were fit following the steps explained in Ref. [20]. Before estimation of the data, it is checked that the compositions at every location meet the condition expressed in Equation 6b to fulfill the condition of Equation 5c. The importance of following this step can be learned in Ref. [21]. From two diffusion couples (DF1 and DF2), as shown in Figure 2, we get four equations with respect to the interdiffusion fluxes of Fe and Co $\left(\widetilde{J}_{Co}\big|_{DF1}, \widetilde{J}_{Fe}\big|_{DF1}, \widetilde{J}_{Co}\big|_{DF2}, \widetilde{J}_{Fe}\big|_{DF2}\right)$ following Equations 7a and 7b (see the supplementary file). After estimating the values of interdiffusion fluxes following Equation 7c, four interdiffusion coefficients are estimated at the composition of the intersection $(Ni_{0.883}Co_{0.03}Fe_{0.037}Mo_{0.05})$ as $\widetilde{D}^{Ni}_{CoCo} = 1.2 \times 10^{-15}$, $\widetilde{D}^{Ni}_{CoFe} = -6.9 \times 10^{-16}$, $\widetilde{D}^{Ni}_{FeFe} = 1.7 \times 10^{-15}$ and $\widetilde{D}^{Ni}_{FeCo} = -2.9 \times 10^{-16}$ $m^2$/s.

It should be noted here that any other component can be considered as the dependent variable as long as the estimated data fulfills the stability equations (see supplementary file) [1,4]. Moreover, data estimated considering different component as dependent variables are related to each other as can be found in reference books [4]. In this analysis, the main and cross interdiffusion coefficients of Mo are not estimated since this component is considered to keep constant. If needed, these diffusion coefficients can be estimated in another set of diffusion couples by keeping any other component instead of Mo as constant.

The importance of this study can be summarized as:

(i) The concept of pseudo ternary approach is proposed by tweaking the experimental method to counter the mathematical complications in a multi component system developed based on Onsager's formalism. As explained in this manuscript, the composition dependent values of the main and cross interdiffusion coefficients

(ii) can be estimated in a system with more than three components, which was not possible earlier based on conventional or alternative available methods.

(ii) In fact, the alternative method of estimating an average value was developed since the composition dependent values of the composition of interest cannot be determined following the conventional method in a multicomponent system [8]. However, these cannot be considered as material constants since the values vary depending on the composition range over which these are estimated or the composition range of the diffusion couples and therefore lack physical significance [5,18]. Keeping in mind a composition of interest for the estimation of the diffusion data, many diffusion couples could be produced by permutation and combination of compositions, as rightly explained by Morral [22]. However, different couples will follow different diffusion (composition) paths. These compositions have their unknown but own composition dependent values of the diffusion coefficients. Therefore, different couples will give different values of the average diffusion coefficients because of different composition range considered for the calculation of these data even if all the diffusion couples pass through the same composition of interest. Additionally, different diffusion couples prepared with different composition ranges follow different diffusion paths even when the tie lines connecting the compositions of the end members overlap [19]. It is known that the diffusion path strongly depends on the end member compositions. This lead to the estimation of different values of the average interdiffusion coefficients in different diffusion couples.

(iii) It might be difficult to produce pseudo ternary profiles in a certain composition range in a material system because of development of (uphill) diffusion profile of the component(s) which are supposed to remain constant throughout the diffusion couple. In the same system, these profiles could be developed successfully by making another component as constant or by selecting a different composition range. Similarly, there might be many other important material systems in which these types of profiles can be developed without much difficulty. The situations are experienced while producing the pseudo binary diffusion couples. Most importantly, irrespective of difficulties, this method gives an opportunity to estimate the composition dependent values of the diffusion coefficients in a multicomponent system, which was not simply possible following the conventional method or any other alternative methods.

(iv) In many multi component materials in applications, two or three major components are present along with few other minor components. The pseudo ternary method gives a possibility of systematic study of diffusional interactions among major components for different concentrations of minor components. With a proper plan, one can even make a systematic study in high entropy multi principal alloys based on the estimation of the composition dependent values of the diffusion coefficients.

(v) In combination with previously proposed pseudo binary approach [11,12] in a multi component system following which one can even estimate the intrinsic diffusion coefficients [12], we can now estimate all types of diffusion coefficients required for understanding the basic phenomenological diffusion mechanisms in correlation with defects and driving forces [13, 16, 17] or examine the diffusional interactions between different components. In combination with the physicochemical approach [4, 23], diffusion controlled microstructural evolution in multicomponent systems can be examined based on the quantitative analysis.

(vi) Most importantly the pseudo ternary method proposed in this manuscript follows relatively easy steps for the estimation of the diffusion coefficients and therefore graduate students and research engineers can easily implement it in their studies. Even the newly proposed numerical methods can be verified based on experimentally measured diffusion coefficients, which was not possible earlier.

**Acknowledgement:** We would like to acknowledge the financial support from ARDB, India.

Figure captions:

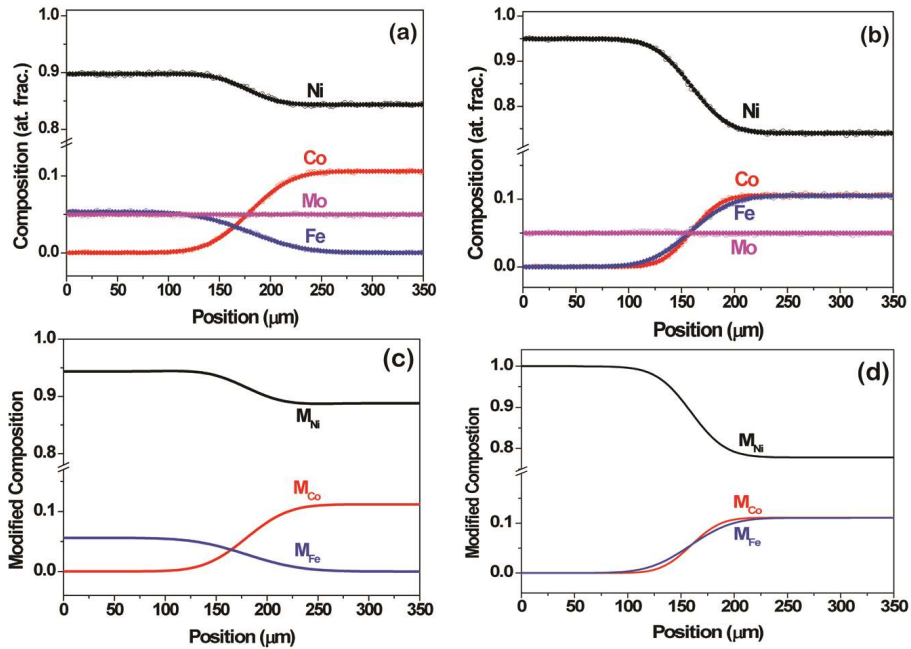

figure 1 (a) Composition and (c) modified composition profiles of $Ni_{90}Fe_5Mo_5/Ni_{85}Co_{10}Mo_5$, (b) composition and (d) modified composition profiles of $Ni_{95}Mo_5/Ni_{75}Co_{10}Fe_{10}Mo_5$ pseudo ternary diffusion couples after annealing at 1100°C for 100h.

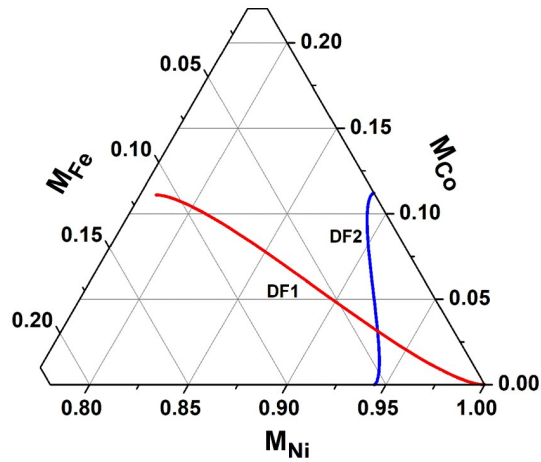

Figure 2 Modified composition profiles on a Gibb's triangle to locate the composition of intersection of pseudo ternary diffusion couples at which the interdiffusion coefficients are estimated.

# Supplementary: Experimental procedure and details of estimation

## 1. Experimental procedure:

The alloys for preparing the diffusion couples were melted using the pure (99.95 wt.%) components in an arc melting furnace. To ensure a better homogeneity, these were remelted five times after flipping the side. These alloys were then homogenized at 1200±5 °C (1473±5 K) for 50 h in high vacuum (~$10^{-4}$ Pa). Following compositions were measured randomly at various spots in an EPMA (Electron Microprobe Analyzer) and the deviation from the average composition was found to be within ±0.1 at.%. These were then EDM (electro discharge machine) cut into ~1 mm thickness, prepared metallographically for flat and smooth surface and coupled in a special fixture (as described in Chapter 3 in Ref. 5). After the isothermal annealing, the diffusion couples were cross-sectioned, prepared metallographically and the composition profiles were measured in EPMA with pure components as standards.

## 2. Estimation of the diffusion coefficients

Total four equations we get from two diffusion couples at the composition of intersection as

$$\widetilde{J}_{Co}\Big|_{DF_1} = -\widetilde{D}^{Ni}_{CoCo} \frac{1}{V_m}\left(\frac{dM_{Co}}{dx}\right)_{DF_1} - \widetilde{D}^{Ni}_{CoFe} \frac{1}{V_m}\left(\frac{dM_{Fe}}{dx}\right)_{DF_1} \tag{S1}$$

$$\widetilde{J}_{Fe}\Big|_{DF_1} = -\widetilde{D}^{Ni}_{FeFe} \frac{1}{V_m}\left(\frac{dM_{Fe}}{dx}\right)_{DF_1} - \widetilde{D}^{Ni}_{FeCo} \frac{1}{V_m}\left(\frac{dM_{Co}}{dx}\right)_{DF_1} \tag{S2}$$

$$\widetilde{J}_{Co}\Big|_{DF_2} = -\widetilde{D}^{Ni}_{CoCo} \frac{1}{V_m}\left(\frac{dM_{Co}}{dx}\right)_{DF_2} - \widetilde{D}^{Ni}_{CoFe} \frac{1}{V_m}\left(\frac{dM_{Fe}}{dx}\right)_{DF_2} \tag{S3}$$

$$\widetilde{J}_{Fe}\Big|_{DF_2} = -\widetilde{D}^{Ni}_{FeFe} \frac{1}{V_m}\left(\frac{dM_{Fe}}{dx}\right)_{DF_2} - \widetilde{D}^{Ni}_{FeCo} \frac{1}{V_m}\left(\frac{dM_{Co}}{dx}\right)_{DF_2} \tag{S4}$$

One can estimate the four interdiffusion coefficients from the above equations after estimating the interdiffusion fluxes following the Equation 7c.

## 3. Stability Equations:

A particular component (k) in a ternary (i,j,k) or pseudo-ternary system can be considered as the dependent variable if the estimated diffusion coefficients fulfil the conditions [1]

$$\widetilde{D}^k_{ii} + \widetilde{D}^k_{jj} > 0 \tag{S5}$$

$$(\widetilde{D}^k_{ii} + \widetilde{D}^k_{jj})^2 \geq 4(\widetilde{D}^k_{ii}\widetilde{D}^k_{jj} - \widetilde{D}^k_{ij}\widetilde{D}^k_{ji}) \tag{S6}$$

$$(\widetilde{D}^k_{ii}\widetilde{D}^k_{jj} - \widetilde{D}^k_{ij}\widetilde{D}^k_{ji}) \geq 0 \tag{S7}$$